\documentclass[aps,preprint,superscriptaddress]{revtex4}      
\usepackage{graphicx}   
\usepackage{subfigure}
\usepackage{float}
\usepackage{amsmath}     
\usepackage{amsfonts}
\usepackage{color}
\usepackage{bm}
\usepackage{bbm}
\usepackage{txfonts}
\usepackage{array}
\usepackage{amssymb}
\usepackage{tabularx}  

\usepackage{hyperref}

\begin{document}

\title{Electron-positron pair production in spatially inhomogeneous electric fields with quadratically symmetric chirp}            
\author{Le-Le Chen}
\affiliation{College of Physics and Electronic Engineering, Northwest Normal University, Lanzhou 730070,China}
\affiliation{School of Mathematics and Physics, Lanzhou JiaoTong University, Lanzhou 730070, China}
\author{Xiao-Ting Xu}
\affiliation{College of Physics and Electronic Engineering, Northwest Normal University, Lanzhou 730070,China}
\affiliation{School of Mathematics and Physics, Lanzhou JiaoTong University, Lanzhou 730070, China}
\author{Rong-An-Tang}
\affiliation{College of Physics and Electronic Engineering, Northwest Normal University, Lanzhou 730070,China}
\author{Xue-Ren Hong \footnote{hongxr@nwnu.edu.cn.}}
\affiliation{College of Physics and Electronic Engineering, Northwest Normal University, Lanzhou 730070, China}
\author{Lie-Juan Li \footnote{ljli@lzjtu.edu.cn.}}
\affiliation{School of Mathematics and Physics, Lanzhou Jiaotong University, Lanzhou 730070, China}
\author{Bai-Song Xie \footnote{bsxie@bnu.edu.cn}}
\affiliation{Key Laboratory of Beam Technology of the Ministry of Education, and School of Physics and Astronomy, Beijing Normal University, Beijing 100875, China}
\affiliation{Institute of Radiation Technology, Beijing Academy of Science and Technology, Beijing 100875, China}
\begin{abstract}   
Electron-positron pair production from vacuum in spatially inhomogeneous electric fields with quadratically symmetric chirp is studied within the real-time Dirac-Heisenberg-Wigner formalism.
The reduced momentum spectrum and the reduced total number of the created particles under the quadratically symmetric chirped electric field are investigated in high- and low-frequency fields.
Compared with that of the quadratically asymmetric chirped field in particular, it is found that the momentum spectrum under the quadratically symmetric chirped field exhibits the stronger oscillations and the higher peaks in both high- and low- frequency fields, and it shows an obvious widening only in the high-frequency field. It is also found that the total number of the created particles of the quadratically symmetric chirped field increases with the chirp, and it is nearly twice that of the quadratically asymmetric chirped field.
\end{abstract}

\maketitle
\section{INTRODUCTION}
Quantum vacuum fluctuations occur when quantum vacuum undergoes significant energy changes over extremely short timescales in strong external fields.
Then, the vacuum becomes unstable and leads to the creation and annihilation of virtual particle pairs,
and they are pulled apart by these fields and no longer annihilate.
At last, they evolve into real particles pairs, for example, electron-positron ($e^{+}e^{-}$) pairs.
This phenomenon is well known as the Schwinger effect, which is one of the long-standing nonperturbative predictions of quantum   electrodynamics (QED)
\cite{sauter1931verhalten,Heisenberg1936,Schwinger1951,DiPiazza2012,Riek2015}.  
However, it has not been observed experimentally because the highest laser intensity that can currently be achieved is
$10^{23}\ \mathrm{W}/\mathrm{cm}^{2}$ \cite{Yoon2021},
and it is far lower than the corresponding critical laser intensity
$I_{cr}\approx4.3\times10^{29}\ \mathrm{W}/\mathrm{cm}^{2}$
of the critical field strength $E_{cr}\approx1.32\times10^{16}\ \mathrm{V}/\mathrm{cm}$,
which is the value that the work done by the electron over a Compton wavelength $\lambda_{c}=386\ \mathrm{fm}$ is equal to the rest energy of the particle,
i.e., $eE_{cr}\lambda_{c}=mc^{2}=0.51\ \mathrm{MeV}$, where $m$ and $e$ are the mass and charge of the electron, and $\hbar$ and $c$
are the reduced Planck constant and the speed of light, respectively \cite{Brezin1970}.
Fortunately, with the rapid development of modern laser technology, especially the invention of chirped pulse amplification (CPA) technique \cite{Strickland1985,Maine1988}, the intensity of laser has been greatly increased by some high-intensity optical laser facilities \cite{Blaschke2006}.
such as the High Power laser Energy Research (HiPER), the Extreme Light Infrastructure (ELI) \cite{Tajima2009}, and the X-ray Free
Electron Laser (XFEL) facilities \cite{Ringwald2001,Alkofer2001}.
Therefore, the field strength can potentially reach one or two orders of magnitude below the critical field strength
\cite{Dumlu2010}, which provides the strong field required for studying the $e^{+}e^{-}$ pair production.

The advancement of modern laser technology is also conducive to researchers exploring significant phenomena of $e^{+}e^{-}$ pair production by various methods.
These research methods include the worldline instantons technique  
\cite{Affleck1982,Dunne2005,Kim2002},
the $S$-matrix method \cite{Hu2010,Ilderton2011},
the semiclassical approach such as the Wentzel-Kramers-Brillouin (WKB) approximation \cite{Strobel2015,Akkermans2012},
the quantum kinetic methods include quantum Vlasov equation (QVE) \cite{Kohlfurst2014},
the Dirac-Heisenberg-Wigner (DHW) formalism
\cite{Kohlfurst2018a,Kohlfurst2020,Li2015,Blinne2016,BlinnePhD2016},
the computational quantum field theory (CQFT) by solving the Dirac equation numerically
\cite{Braun1999,Gerry2006,Jiang2012,Xie2017},
as well as the quantum two level model (TLM)
\cite{DiPiazza2001,Fiordilino2021, Amat2025,Amat2025b}, and so on.
Previous investigations have used several forms of external field configurations to study the $e^{+}e^{-}$ pair production through the above methods,
such as the constant fields \cite{Narozhnyi1976},
time-dependent electric fields
\cite{Roberts2002,Hebenstreit2009,Dumlu2011,Hu2023},
time delay electric fields \cite{Kullie2024},
spacetime-dependent electric fields \cite{Kim2007,Kleinert2008,Hebenstreit2011,Hu2023CTP},
the electric fields of dynamically assisted Schwinger mechanism that combines two laser pulses with a low-frequency strong field and a high-frequency weak field
\cite{Orthaber2011,Schutzhold2008,DiPiazza2009},
electromagnetic fields
\cite{BialynickiBirula1991,Kim2006,Bulanov2010,Tanji2009}, and
polarization electric fields \cite{He2012,Aleksandrov2024,Tang2022}.
The general mechanisms behind the pair creation process,
such as interference effects \cite{Kohlfurst2018b},
particle self-bunching effects \cite{Dumlu2011},
ponderomotive effects \cite{Kohlfurst2018PRD}, and
spin effect \cite{Chen2025},
can be successively discovered from different external configurations.
Particularly, recent research has indicated that the production of $e^{+}e^{-}$ pairs in different chirped electric fields is highly significant.
For example,
the effects of the chirp on the momentum spectrum of the produced particles in the chirped laser pulses with a subcycle structure are analyzed, and the result shows that chirp parameters are vital for the momentum spectrum of the produced particles
\cite{Dumlu2010a}.
The momentum spectrum of the produced $e^{+}e^{-}$ pairs for several linear frequency chirps
and the sensitivity of the particle number density to chirp variations in the homogeneous single-pulse field are investigated, and
it reveals that the linear frequency chirps lead to the strong interference effects,
resulting in the substantial oscillations in the momentum spectra, and the particle number density is very sensitive to the variations of the chirp parameter \cite{Olugh2019}.
The effects of chirp on $e^{+}e^{-}$ pairs creation in the space-time dependent electric fields are studied, and
it finds that the total particle number enhancement only occurs for larger chirp in the rapidly oscillating field,
while in the slowly oscillating field, the chirp have crucial effects on the $e^{+}e^{-}$ pairs momentum distribution
\cite{Ababekri2020}.
The effect of external field parameters on the momentum spectrum and the total number of the produced $e^{+}e^{-}$ pairs in inhomogeneous electric fields with linear symmetric frequency chirp is studied, and
the results show that the momentum spectrum is sensitive to the spatial scales and the carrier phase, and the total particle number
increases with increasing chirp \cite{Mohamedsedik2021}.
Recently, the effects of a quadratically asymmetric chirp on $e^{+}e^{-}$ pairs production in spatially inhomogeneous chirped fields are also investigated, and
the results reveal that the quadratically asymmetric chirp influences the momentum spectrum of the generated particles, resulting in a pronounced oscillation of the momentum spectrum \cite{Osman2023}.
However, there has not been sufficient research on the quadratically symmetric chirp in spatially inhomogeneous electric fields so far.

In this paper, the generation of $e^{+}e^{-}$ pairs is studied in spatially inhomogeneous electric fields with quadratically symmetric chirp using the DHW formalism.
The reduced momentum spectrum and the reduced total number of the created particles under the quadratically symmetric chirped field are studied in high- and low-frequency fields.
In particular, they are compared with the results of the quadratically asymmetric chirped field.
It is found that the momentum spectrum under the quadratically symmetric chirped field exhibit the stronger oscillations and the higher peaks in both high- and low- frequency fields,
and it shows a significant widening only in the high-frequency field.
It is also found that the total number of the created particles under the quadratically symmetric chirped field increases as the chirp increases, and it is nearly twice that of the quadratically asymmetric chirped field.
Note that the natural units, where $\hbar=c=1$ are applied and all quantities are express in terms of the electron mass $m$
and the electron Compton wavelength $\lambda_{c}$, where $\lambda_{c}=m^{-1}$.

This paper is structured as follows.
In Sec. \ref{section2}, the external electric field model is introduced and the DHW formalism is reviewed.
In Sec. \ref{section3}, the momentum spectrum and the total number of the created particles for high- and low-frequency fields are presented.
In Sec. \ref{section4}, the turning-point structure is shown for analyzing the features of the momentum spectrum.
In Sec. \ref{section5}, the conclusion and outlook are given.

\section{EXTERNAL FIELD MODEL AND DHW FORMALISM}     \label{section2}  
\subsection{Model of external field}   

In order to study the generation of vacuum $e^{+}e^{-}$ pairs,
we propose the external field model of spatially inhomogeneous electric field with the quadratically symmetric chirp, which can be described  as \cite{Osman2023}
\begin{equation}
 \begin{aligned}
E(x,t)& = E_0 f(x) g(t)\\& =\epsilon E_{cr}
\exp\left(-\frac{x^2}{2\lambda^2}\right)\exp\left(-\frac{t^{2}}{2\tau^{2}}\right)
\cos\left(\omega t + b_1 |t| t + b_2 |t|^{2}t +\phi\right),
   \end{aligned}
\label{eq:1}
\end{equation}
where $E_0$ represents the field strength, $f(x)$ and $g(t)$ represent the spatial- and time-dependent parts, respectively. The parameter \(\epsilon=0.5\) is set, $\lambda$ and $\tau$ denote the spatial and temporal scales of the external field, $\omega$ and $\phi$ are the original frequency and the carrier-envelope phase, respectively.  
The parameters $b_{1}$ and $b_{2}$ are the chirped parameters and can be represented as \(b_{1}=\alpha\omega/2\tau\), \(b_{2}=\alpha\omega/2\tau^{2}\) $(0\leq\alpha \leq 1)$ in the interaction time interval \(-\tau \leq t\leq \tau\), where $e^{+}e^{-}$ pair creation primarily occurs the peak of the external field and its vicinity.
Here, the difference between the quadratically symmetric chirped
electric field of the Eq. (\ref{eq:1}) and the quadratically asymmetric chirped electric field of the Ref. \cite{Osman2023} is the effective frequency $\omega_{eff}$.
For the quadratically symmetric chirped electric field, $\omega_{eff}(t)=\omega + b_{1} |t|+ b_{2} |t|^2$,
while for the quadratically asymmetric chirped electric field, $\omega_{eff}(t)=\omega + b_{1} t +b_{2}t^2$,
and their upper limit can be set as $\omega_{eff}\leq 2\omega$.

In the following, only the result of $\phi=\pi/2$ is presented for brevity.
This is because in the low-frequency field for the case of $\phi=0$,
it is found that the total number of generated $e^{+}e^{-}$ in the quadratically symmetric chirped electric field is basically the same with the quadratically asymmetric chirped electric field.
However, for the case of $\phi=\pi/2$, it shows a significant increase in the quadratically symmetric chirped electric field. Here, it investigates two fields: the high and low-frequency field.
For the high-frequency field, the parameters are set as $\omega=0.7\ \mathrm{m}$ and $\tau=45\ \mathrm{m}^{-1}$, and for the low-frequency field, the parameters are set as $\omega=0.1\ \mathrm{m}$ and $\tau=25\ \mathrm{m}^{-1}$. Note that the parameters taken here are the same as those in Ref. \cite{Ababekri2020}.

This electric field model is a standing wave formed by two counter propagating coherent lasers with spatial dependence.
Since the lasers propagating in the opposite directions cancel out the magnetic field component, and the most of the particles at the laser focus point are generated along the electric field direction,
thus this model can be simplified to 1+1 dimensions setting by ignoring the particle momenta perpendicular to this dominant direction
\cite{HebenstreitPhD2011}.
The electric field direction is along the $x$-axis, and the field strength varies with both $x$ and $t$.

\subsection{DHW formalism}  

The DHW method is the density operator formed by the commutation of two creation and annihilation operators in the Heisenberg picture
\cite{Heisenberg1925}.
It is a quantum kinetic method in phase space for studying $e^{+}e^{-}$ pair production under both spatially homogeneous
\cite{Blinne2014,Blinne2016} and inhomogeneous
electromagnetic fields \cite{Kohlfurst2016}.
The complete calculation of the DHW formalism can be found in Refs.
\cite{Kohlfurst2018PRD,HebenstreitPhD2011,Ababekri2019}.
In this paper, the relevant equations and key concepts are briefly introduced.

To study $e^{+}e^{-}$ pair production in electromagnetic fields, it starts with the QED Lagrangian $\mathcal{L}$, and it is given by
\begin{equation}
\mathcal{L}(\hat{\Psi},\bar{\hat{\Psi}}, \hat{\vec{A}})= \frac{1}{2} (i \bar{\hat{\Psi}} \gamma^{\mu} \mathcal{D}_{\mu}\hat{\Psi} - i \bar{\hat{\Psi}} \mathcal{D}_{\mu}^{\dagger} \gamma^{\mu} \hat{\Psi})-m \bar{\hat{\Psi}} \hat{\Psi}-\frac{1}{4} F^{\mu \nu}F_{\mu \nu},
\end{equation}
where $\hat{\Psi}$ is the Dirac field operator, $\bar{\hat{\Psi}}$ is the conjugate field operator, $\hat{\vec{A}}$ is the  four-vector-potential of the electromagnetic field,
$F^{\mu \nu}=\partial^{\mu}A^{\nu}-\partial^{\nu}A^{\mu}$ and $F_{\mu \nu}=\partial_{\mu}A_{\nu}-\partial_{\nu}A_{\mu}$ are the field strength tensor, $\mathcal{D}_{\mu}=( \partial_{\mu}+ i e A_{\mu})$ and $\mathcal{D}_{\mu}^{\dagger}=( \overset{\leftharpoonup}{\partial}_{\mu} - i e A_{\mu})$ denote the covariant derivatives, $A_{\mu}$ is electromagnetic field, $\gamma^{\mu}$ represents the gamma matrices.

To obtain the equations of motion for the particle, the Euler-Lagrange equation is given by  \cite{Serfaty2014}
\begin{equation}
\partial_{\mu} \left[ \frac{ \partial \mathcal{L}}{\partial (\partial_{\mu} \bar{\hat{\Psi}})} \right]= \frac {\partial \mathcal{L}}  {\partial \bar{\hat{\Psi}}}.
\label{eq:po}
\end{equation}

By applying the variational principle to the Euler-Lagrange equation of Eq. (\ref{eq:po}), the Dirac equation of motion for the particle is obtained as \cite{Dirac1928a}
\begin{equation}
\left[ i\gamma^{\mu} \partial_{\mu}-e\gamma^{\mu}A_{\mu}(r)-m \right]\Psi(r)=0,
\label{eq:a}
\end{equation}
where $\Psi(r)$ is the Dirac wave function, $\partial_{\mu}$ is four-dimensional partial derivative.

In order to describe the creation and annihilation of $e^{+}e^{-}$ pair in the vacuum, the fermion field operator $\hat{\Psi}_{\alpha}$ (creation operator) and its conjugate operator ${\hat{\Psi}^{\dag}}_{\beta}$ (annihilation operator) can be defined, and they construct an equal-time density operator $\hat{\mathcal{C}}_{\alpha\beta}^{\pm}$ formed by commutators of two field operators in the Heisenberg picture as 
\begin{equation}
\hat{\mathcal{C}}_{\alpha\beta}^{\pm}(t,\mathit{r}, \mathit{s}) = \langle 0 | \hat{\Psi}_{\alpha}(t, r) \hat{\Psi}_{\beta}^{\dagger}(t, s) | 0 \rangle \pm  \langle 0 | \hat{\Psi}_{\beta}^{\dagger}(t, s) \hat{\Psi}_{\alpha}(t, r) | 0 \rangle,
\end{equation}
where $r=(r_{1}+r_{2})/2$ is the center-of-mass coordinate and $s=r_{2}-r_{1}$ is the relative coordinate,
and it is important to note that what is used here is the adjoint field operator $\bar{\hat{\Psi}}_{\beta}= {\hat{\Psi}^{\dag}}\gamma^{0}$.    
Then, the covariant density operator $\hat{\mathcal{C}\hspace{0.001em}}_{\alpha\beta}$ of the system is defined as
\cite{BialynickiBirula1991}
\begin{equation}
\hat{\mathcal{C}\hspace{0.001em}}_{\alpha\beta}(\mathit{r},\mathit{s})= \mathcal{U}(A, r, s)\left[\bar{\hat{\Psi}}_{\beta}(\mathit{r}-\frac{s}{2}),\Psi_{\alpha}(r+\frac{s}{2})\right],
\label{eq:4}  
\end{equation}
where $\mathcal{U}\left(\mathit{A}, \mathit{r}, \mathit{s}\right)$ is the Wilson line factor introduced to ensure gauge invariance and given by
\begin{equation}
\mathcal{U}(\mathit{A}, \mathit{r}, \mathit{s}) = \exp \left[ i e s \int_{-\frac{1}{2}}^{\frac{1}{2}} \mathrm{d}\xi \, A \left(r+\xi s\right)\right].
\end{equation}
Here, it is seen that $\mathcal{U}(\mathit{A}, \mathit{r}, \mathit{s})$ depends on the elementary charge $e$ and the background gauge field $A$.
Treating the background gauge field in the mean-field (Hartree) approximation, i.e.,
\begin{equation}
F^{\mu\nu}(x)\approx \langle \Phi| \hat{F}^{\mu\nu}(x)|\Phi\rangle,
\end{equation}
it means the electromagnetic field strength tensor $\hat{F}^{\mu\nu}$ is treated as a C-number valued function, and $\Phi$ is vacuum state.
This approximation becomes apparent when we consider terms of the form $\langle \hat{F}^{\mu\nu}(x)   \hat{\mathcal{C}}_{\alpha\beta}(r,s)\rangle$, which can be expressed as
\cite{Hebenstreit2011,BialynickiBirula1991,HebenstreitPhD2011,Vasak1987}

\begin{equation}
\langle \Phi|  \hat{F}^{\mu\nu}(x) \hat{\mathcal{C}}_{\alpha\beta}(r,s) | \Phi  \rangle \approx F^{\mu\nu}(x) \langle \Phi|  \hat{\mathcal{C}}_{\alpha\beta}(r,s)| \Phi \rangle.
\end{equation}
After performing the Fourier transform on Eq. (\ref{eq:4}), the covariant Wigner operator $\hat{\mathcal{W}}_{\alpha\beta}$ is  obtained as
\begin{equation}
\hat{\mathcal{W}}_{\alpha\beta}(\mathit{r},\mathit{p})= \displaystyle\frac{1}{2}\int \mathrm{d}^4 s\, \exp(ips)\hat{\mathcal{C}\hspace{0.1em}}_{\alpha \beta} (\mathit{r},\mathit{s}).
\label{eq:m}
\end{equation}
By taking the vacuum expectation value of the covariant Wigner operator $\hat{\mathcal{W}}_{\alpha\beta}$, the covariant Wigner function $\mathbb{W}(\mathit{r}, \mathit{p})$ can be obtained as
\begin{equation}
\mathbb{W} (\mathit{r}, \mathit{p})= \langle \Phi | \hat{\mathcal{W}}_{\alpha\beta}(\mathit{r}, \mathit{p}) | \Phi \rangle.
\label{eq:b}
\end{equation}

Then, the $\mathbb{W}(\mathit{r}, \mathit{p})$ is decomposed using a complete basis set
$\left\{\mathbb{I},\gamma_5,\gamma^{\mu},\gamma_5 \gamma^{\mu},\sigma^{\mu\nu}:=\frac{i}{2}[\gamma^{\mu},\gamma^{\nu}]\right\}$ into 16 covariant Wigner components in 3+1 dimensions as \cite{KohlfurstPhD2015}
\begin{equation}
\mathbb{W}=\displaystyle\frac{1}{4}(\mathbb{IS}+i \gamma_5 \mathbb{P}+\gamma^\mu \mathbb{V}_\mu+\gamma_5 \gamma^\mu \mathbb{A}_\mu + \sigma^{\mu\mathit{\nu}} \mathbb{T}_{\mu\nu}),
\label{eq:8}
\end{equation}
where $\mathbb{S}$, $\mathbb{P}$, $\mathbb{V}_{\mu}$, $\mathbb{A}_{\mu}$ and $\mathbb{T}_{\mu\nu}$ represent as scalar,   pseudoscalar, vector, axial vector and tensor, respectively, with $\mu=0,1,2,3$.
Considering that the form of external vector potential satisfies $\mathbf{A}=A(x,t) \mathbf e_{x}$, so Eq. (\ref{eq:8}) can be simplified to the case of 1+1 dimensions as
\begin{equation}
\mathbbm{W} = \displaystyle\frac{1}{2} (\mathbb{IS} + i \gamma_5 \mathbb{P} + \gamma^\mu \mathbb{V}_\mu),
\label{eq:c}
\end{equation}
with $\mu=0,1$.

It is note that the motion equation of the covariant Wigner operator $\hat{\mathcal{W}}_{\alpha\beta}$ can be derived from Eqs. (\ref{eq:a}), (\ref{eq:4}) and (\ref{eq:m}), Then, the equations of motion for the covariant Wigner function $\mathbb{W}(\mathit{r}, \mathit{p})$ can be obtained from Eq. \ref{eq:b} as
\begin{align}
\left[ \frac{1}{2} D_{\mu}(r,p)+i \Pi_{\mu}(r,p)\right] \mathbbm{W}(r,p)\gamma^\mu =i\mathbbm{W}(r,p), \label{eq:e}\\
\left[ \frac{1}{2} D_{\mu}(r,p)-i\Pi_{\mu}(r,p)\right] \gamma^\mu \mathbbm{W}(r,p)=-i\mathbbm{W}(r,p),
\label{eq:f}
\end{align}
where the derivative operators are given by
\begin{align}
D_{\mu}(r,p) = \partial_{\mu}^{r}-e\int \mathrm{d}\xi \, F_{\mu\nu} \left( r - i\xi \partial_{p} \right) \partial_{p}^{\nu}, \\
\Pi_{\mu}(r,p) =p_{\mu} - ie \int \mathrm{d}\xi \, \xi F_{\mu\nu} \left( r - i \xi \partial_{p} \right) \partial_{p}^{\nu}.
\end{align}

By substituting the expansion from Eq. (\ref{eq:c}) into the equations of motion for the covariant Wigner functions of Eqs.      (\ref{eq:e}) and (\ref{eq:f}), and by considering the commutator and anti-commutator relations of all Dirac bilinears in 1+1 dimensions with $\gamma^\mu$ matrices,
a system of partial differential equations (PDEs) that includes all covariant Wigner components can be determined.

To express Eq. (\ref{eq:c}) as an initial value problem, the equal-time Wigner function is defined as the energy average of the covariant Wigner function \cite{Blinne2014}, and it can be be written as
\begin{equation}
\mathbbm{w}(\mathbf{x}, \mathbf{p},t)=\int \frac {\mathrm{d} p_0} {2\pi}\mathbb{W}(r,p),
\end{equation}
where $\mathbf{x}$ and $\mathbf{p}$ represent the position of the particle and the kinetic momentum, respectively. Therefore, the equations of motion for the Wigner function components in 1+1 dimensions are described as \cite{KohlfurstPhD2015}
\begin{align}
D_t\mathbbm{s} - 2p_x \mathbbm{p} = 0,     \label{eq:11} \\
D_t\mathbbm{v}_0 + \partial_x \mathbbm{v}_1 = 0,  \label{eq:12}\\
D_t\mathbbm{v}_1+\partial_x \mathbbm{v}_0=-2\mathbbm{p}, \label{eq:13}\\
D_t\mathbbm{p}+ 2 p_x \mathbbm{s}=2\mathbbm{v}_1.    \label{eq:14}
\end{align}
with the pseudodifferential operator
\begin{equation}
D_t=\partial t + \int_{-\frac{1}{2}}^{\frac{1}{2}}\mathrm{d}\xi \, E_x  ( x+i \xi \partial_{p_x},t) \partial{p_x}.
\end{equation}
The vacuum initial condition values are applied to the system by \cite{Ababekri2020}
\begin{equation}
\mathbbm{s}_{\mathrm{vac}}=-\frac{2m} {\Omega(p_{x})},\hspace{1cm}\mathbbm{v}_{1\mathrm{vac}}=-\frac{2p_x}{\Omega(p_{x})},
\label{eq:16}
\end{equation}
where $\Omega$ represents the energy of individual particles, which can be expressed as $\Omega =\sqrt{p_x^2+m^2}$. Subtracting the vacuum expectation value in Eq. (\ref{eq:16}) is equivalent to normalizing of the Dirac sea as
\begin{equation}
\mathbbm{w}^{\upsilon}_{k} (x,p_{x},t)= \mathbbm{w}_{k}(x,p_{x},t)-\mathbbm{w}_{\mathrm{vac}}(p_{x}),
\end{equation}
where $\mathbbm{w}_{k}$ is the Wigner component in Eqs. (\ref{eq:11})-(\ref{eq:14}). For the corresponding symbol, it is represented as $\mathbbm{w}_{0}=\mathbbm{s}$, $\mathbbm{w}_{1}=\mathbbm{v}_{0}$, $\mathbbm{w}_{2}=\mathbbm{v}_{1}$,     $\mathbbm{w}_{3}=\mathbbm{p}$, and the $\mathbbm{w}_{\mathrm{vac}}$ denotes the corresponding vacuum initial condition.

Finally, it is known that the energy of the particle yield is $m\mathbbm{s}^{\upsilon}(x,p_{x},t)+p_{x} \mathbbm{v}^{\upsilon}_{1}$, which leads to the expression for the particle number density in phase space as
\begin{equation}
n(x, p_x, t)=\frac{m \mathbbm{s}^{\upsilon}(x,p_x,t)+ p_x \mathbbm{v}^{\upsilon}_{1} (x,p_x,t)}{\Omega (p_x)}
\label{eq:18}.
\end{equation}
Then, the momentum distribution function of the created particles could be obtained by integration of Eq. (\ref{eq:18}) with respect to $x$ as
\begin{equation}
n(p_x, t)=\int \mathrm{d}x\,n(x,p_x,t),
\end{equation}
and the total particle number is obtained by integrating the whole phase space as
\begin{equation}
N(t)= \int \frac {\mathrm{d}p_x}{2\pi}n(p_x, t).
\label{eq:20}
\end{equation}

In order to extract the nontrivial effect of the spatial scale $\lambda$, the reduced quantities $\bar{n} \left(p_x,t\right) = n (p_{x},t)/\lambda$ and $\bar{N}(t\rightarrow\infty)= N(t\rightarrow\infty)/\lambda$ are often used \cite{Hebenstreit2011}.   

\section{NUMERICAL RESULTS}   \label{section3}

In this section, it mainly studies the momentum spectrum and the total number of the created particles with different quadratically  symmetric chirp at various spatial scales in high- and low-frequency fields, respectively.
Particularly, the momentum spectrum and the total particle number for the quadratically symmetric chirp field and the quadratically asymmetric chirped field are compared.
Note that the momentum spectrum for $\lambda=300\ \mathrm{m}^{-1}$ in our study is the same as that $\lambda=1000\  \mathrm{m}^{-1}$ in Ref. \cite{Ababekri2020}, indicating that quasihomogeneous field approximation is already valid at $\lambda=300\   \mathrm{m}^{-1}$.
Therefore, we do not present the momentum spectrum more than $\lambda=300\ \mathrm{m}^{-1}$.
For comparison, the total particle number of the quadratically asymmetric chirped field is presented again, and this result is consistent with the findings of the Ref. \cite{Osman2023}.

Generally, $e^{+}e^{-}$ pairs in a vacuum can be generated through two mechanisms,
which are multiphoton absorption and the tunneling process.
These two mechanisms can be well distinguished by the Keldysh adiabaticity parameter $\zeta=\frac{m\omega}{|e| E_{0}}$
\cite{Keldysh1965,Fedorov2016}.
For the multiphoton absorption $\zeta\gg1$, while for the tunneling process $\zeta\ll1$.
In this paper, the Keldysh parameter $\zeta$ can be derived from the given parameters in Sec. \ref{section2},                                                          and it is $\zeta=1.4$ for the high-frequency field and $\zeta=0.2$ for the low-frequency field.
That is, $\zeta\sim \mathcal{O}(1)$ in the high-frequency field, while $\zeta\ll1$ in the low-frequency field.
For the high-frequency field, although $\zeta\sim \mathcal{O}(1)$ indicates the existence of both mechanisms, it should be noted that multiphoton absorption plays the dominant role according to the existing research findings and our subsequent investigations
\cite{Xie2017,Orthaber2011}.
For the low-frequency field, since $\zeta=0.2\ll1$, it is dominated by tunneling processes.

\subsection{High frequency field}
\subsubsection{Momentum distribution of created particles}


\begin{figure}[htbp]
	\centering
	\includegraphics[width=1\textwidth,height=0.25\textheight]{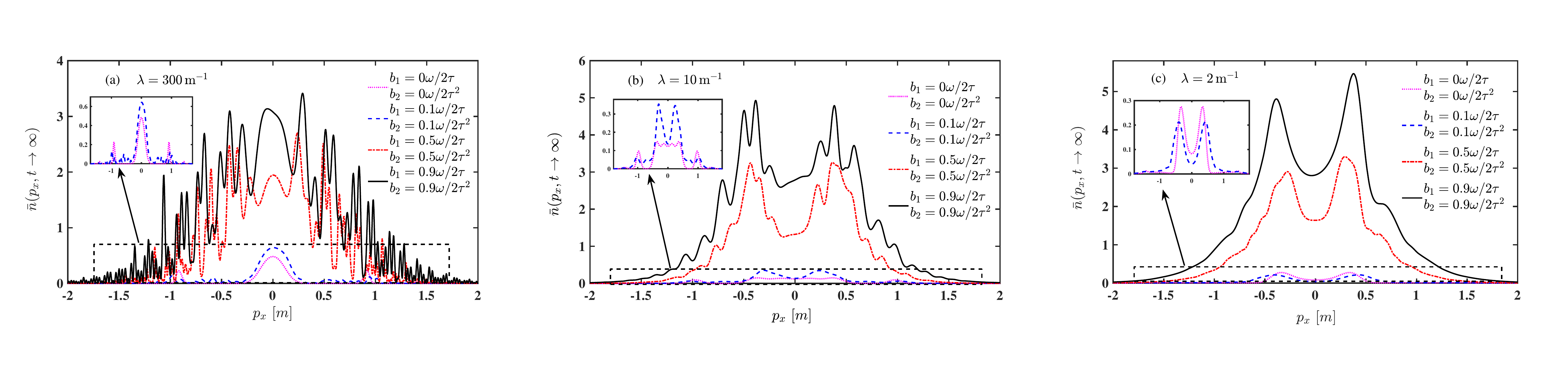}  
	\caption{Reduced momentum spectrum with different quadratically symmetric chirp for various spatial scales in the high-frequency field: (a), (b), and (c) represent the cases where $\lambda=300\ \mathrm{m}^{-1}$, $\lambda=10\ \mathrm{m}^{-1}$, and $\lambda=2\ \mathrm{m}^{-1}$, respectively.
		The insets show the magnified momentum spectrum for the chirp parameters $b_{1}=b_{2}=0$, $b_{1}=0.1\omega/2\tau$ and $b_{2}=0.1\omega/2\tau^{2}$ for clarity.
		The field parameters are $\epsilon=0.5$, $\omega=0.7\ \mathrm{m}$, $\tau=45\ \mathrm{m}^{-1}$, and $\phi=\pi/2$.}
	\label{fig:1}          
\end{figure}        

As can be seen from Fig. \ref{fig:1},      
the overall momentum spectrum shows obvious widening as the chirp increases.
When the spatial scale is quasihomogeneous and there is no chirp, the momentum spectrum exhibits a standard multiphoton peak structure and shows symmetry.
However, the introduction of the quadratically symmetric chirp breaks the symmetry and causes oscillations in the momentum spectrum,   which become more complex as the chirp increases.
At spatial scale $\lambda=10\ \mathrm{m}^{-1}$, the momentum spectrum shows a peak-valley-peak structure for small, moderate, and large chirp, and oscillations are present for moderate and large chirp.
However, at an extremely small spatial scale, an obvious peak-valley-peak structure appears in the momentum spectrum among all chirp, and there are no obvious oscillations in the momentum spectrum.

As can be seen from Fig. \ref{fig:1}(a),      
when the spatial scale is $\lambda=300\ \mathrm{m}^{-1}$.  
For the zero chirp, i.e., $b_{1}=b_{2}=0$,    
the momentum spectrum shows symmetric at $p=0$, and multiple peaks are clearly observed.  
These peaks of the momentum spectrum at $p=0$, $p=\pm0.9\ \mathrm{m}$, and $p=\pm1.4\ \mathrm{m}$ correspond to 3, 4 and 5 photon  
absorption, respectively \cite{Kohlfurst2018b},
and these $n$-photon peaks can be interpreted as a result that an interference pattern observable formed by the superposition of
particle trajectories \cite{Nousch2016}.
For the small chirp, i.e., $b_{1}=0.1\omega/2\tau$ and $b_{2}=0.1\omega/2\tau^{2}$,
the weak oscillation appears on both sides of the peak at $p=0$.
When the chirp is moderate, i.e., $b_{1}=0.5\omega/2\tau$ and $b_{2}=0.5\omega/2\tau^{2}$,
the symmetry of the momentum spectrum is broken, and the stronger oscillation appear on the momentum spectrum.
When the chirp is large, i.e., $b_{1}=0.9\omega/2\tau$ and $b_{2}=0.9\omega/2\tau^{2}$,
the intense and complicated oscillation appear on the momentum spectrum.
The characteristics of these oscillations can be explained starting from $\omega_{eff}(t)=\omega + b_{1}|t|+b_{2}|t|^2$.
It is seen that the effective frequency increases significantly with increasing chirp, which supplies more energy and generates a   greater number of $e^{+}e^{-}$ pairs.
As a result, the interactions between these pairs lead to more pronounced interference effects in the momentum spectrum.

When spatial scale decreases to $\lambda=10\ \mathrm{m}^{-1}$, the corresponding momentum spectrum is shown in Fig.  \ref{fig:1}(b).  
For the zero chirp,
the main peak at $p=0$ is significantly broadened compared with the case of $\lambda=300\ \mathrm{m}^{-1}$ here,
while the peaks at $p=\pm0.9\ \mathrm{m}$ and $p=\pm1.4\ \mathrm{m}$ remain visible, and the momentum spectrum shows symmetry at $p=0$.
The main peak in the momentum spectrum takes on a much broader form which can be related to quantum interferences \cite{Kohlfurst2018b}.  
For the small chirp, this broadening around $p=0$ shows a significant change, forming a peak-valley-peak structure
with peaks at $p=-0.3\ \mathrm{m}$ and $p=0.25\ \mathrm{m}$ and a valley at $p=0$,
this indicates the broken symmetry of the momentum spectrum.
This peak-valley-peak structure appears from the ponderomotive force induced by spatial inhomogeneity, as well as the additional energy provided by the small chirp introduced into the external field, which causes the particles to spread out toward both sides.
For the moderate chirp, the peaks of the momentum spectrum are replaced by a weak oscillation.
The strong oscillation is observed in the momentum spectrum at large chirp.

At the extremely small spatial scale $\lambda=2\ \mathrm{m}^{-1}$, the corresponding momentum spectrum is shown in Fig. \ref{fig:1}(c).      
For the zero chirp, it is seen that the momentum spectrum shows two same peaks at $p=\pm0.3\ \mathrm{m}$ and one valley at $p=0$.
Due to the symmetry of the applied electric field in $\pm x$ directions,
the ponderomotive forces push the pairs apart in $+x$ and $-x$ directions, thereby causing the pairs to accelerate in both directions, which results in two same peaks,
and the valley at $p=0$ results from the particle pairs accelerating in the positive or negative directions cancel each other out.
For the small chirp, it is observed that the momentum increases when the peaks appear, with two peaks observed at $p=\pm0.4\ \mathrm{m}$.  
For the moderate and large chirp, the symmetry of the momentum spectrum is broken, and the peaks of the momentum spectrum become higher and the valley deeper.
Moreover, the momentum spectrum shows no significant oscillations.
This is because the spatial inhomogeneity increases as the spatial scales decrease, which results in stronger ponderomotive forces.
It pushes more pairs out of the strong field region, resulting in the weakening of the interference effects between pairs.

\begin{figure}[htbp]  
	\centering
	\includegraphics[width=0.8\textwidth,height=0.4\textheight]{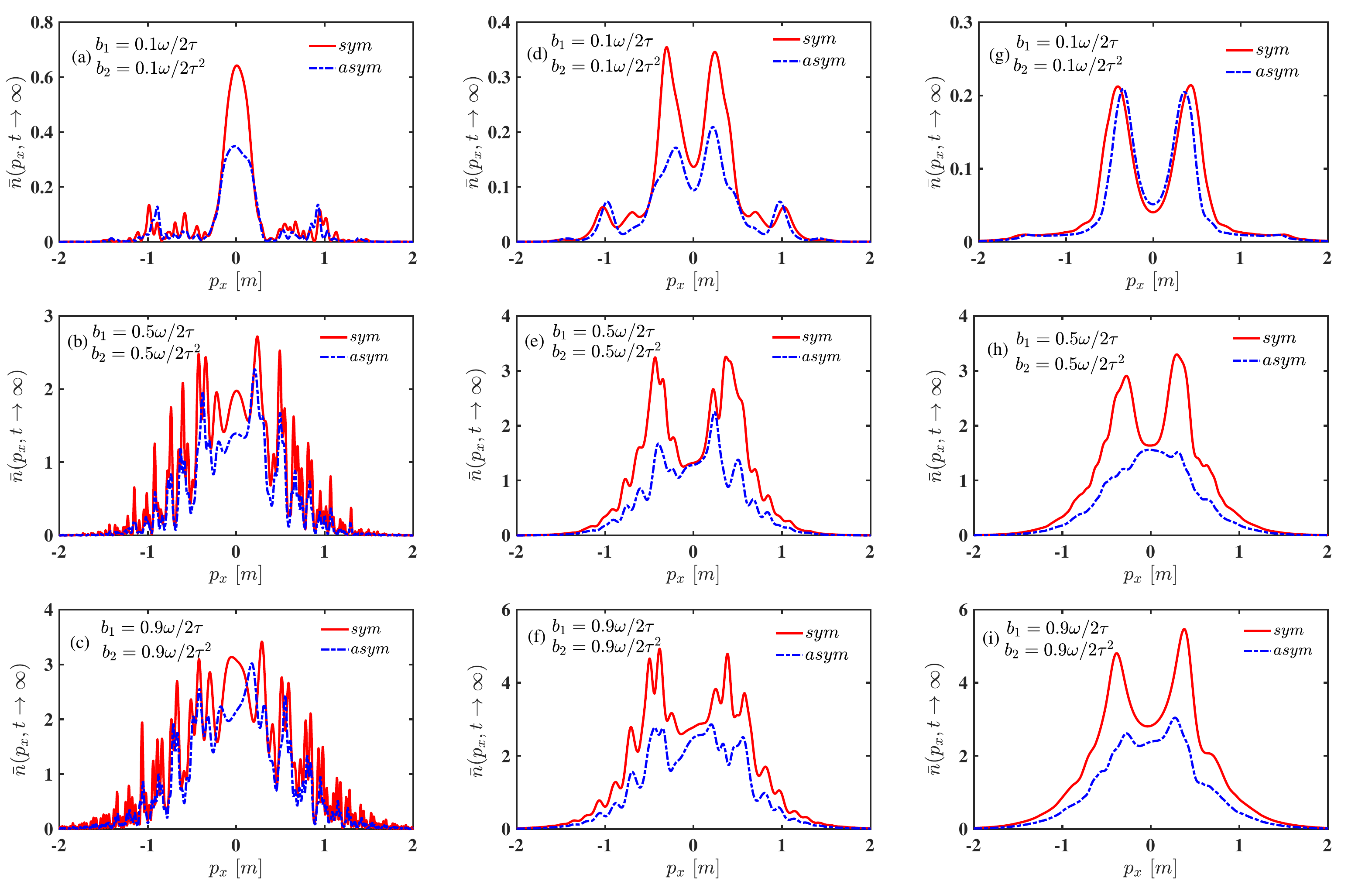}     
	\caption{Comparison of the reduced momentum spectrum with quadratically symmetric and quadratically asymmetric chirp in the    high-frequency field.
		From left to right, the columns (a)--(c), (d)--(f), and (g)--(i) represent the cases of $\lambda=300\ \mathrm{m}^{-1}$, $\lambda=10\ \mathrm{m}^{-1}$, and $\lambda=2\ \mathrm{m}^{-1}$, respectively.
		The field parameters are the same as in Fig. \ref{fig:1}.}    
	\label{fig:2}          
\end{figure}

Furthermore, a comparison of the momentum spectrum for quadratically symmetric and quadratically asymmetric chirped fields is shown in Fig. \ref{fig:2}.                  

It can be clearly seen that the momentum spectrum for the quadratically symmetric chirp exhibits stronger oscillations, as well as higher peak values and obvious broadening.
The reason for the stronger oscillations is that the external time-dependence electric field of the quadratically symmetric chirp vibrates more times in the $-t$ direction than that of the quadratically asymmetric chirp, which leads to a higher frequency.   
As a result, more energy is provided, and more $e^{+}e^{-}$ pairs are produced through the multiphoton absorption process, which causes stronger interference effects due to their interactions.
Meanwhile, the higher peak values and obvious broadening are observed in the quadratically symmetric chirped field due to the increase in energy.  
Moreover, at $\lambda=10\ \mathrm{m}^{-1}$ and $\lambda=2\ \mathrm{m}^{-1}$,
the peak-valley-peak structure of the momentum spectrum under the quadratically symmetric chirp field is more pronounced than that under the quadratically asymmetric one, which is possibly caused by the stronger ponderomotive force in the quadratically symmetric chirp field than in the quadratically asymmetric one.

\subsubsection{Total number of created particles}

\begin{figure}[htbp]
	\centering
	\includegraphics[width=0.8\textwidth,height=0.25\textheight]{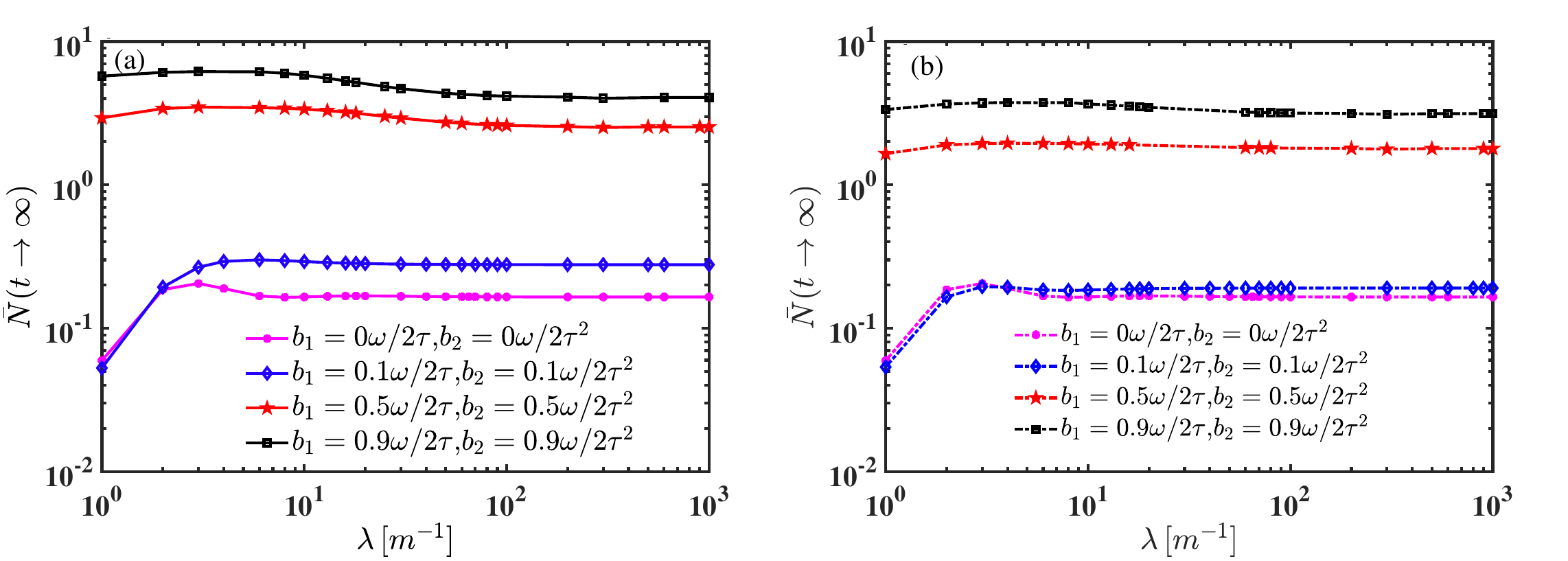} 
	\caption{The variation of the total number of created particles with the spatial scales for different high-frequency quadratically symmetric and quadratically asymmetric chirped fields,
		(a) and (b) represent the cases of the quadratically symmetric chirp (solid line) and quadratically asymmetric chirp (dot-dashed line), respectively.
		The field parameters are the same as in Fig. \ref{fig:1}.}  
	\label{fig:3}      
\end{figure}

The variation of the total number of created particles with spatial scales for different quadratically symmetric and quadratically asymmetric chirp in the high-frequency field is shown in Fig. \ref{fig:3}.    
Note that Fig. \ref{fig:3}(b) is the same as that of Fig. 6(a) in Ref. \cite{Osman2023}.          

It is observed that the stable total number of created particles increases with increasing chirp in both quadratically symmetric and asymmetric chirped fields.
Furthermore, when $b_{1}=b_{2}=0$, and $b_{1}=0.1\omega/2\tau$ and $b_{2}=0.1\omega/2\tau^{2}$,
the total particle number show a significant increase at smaller spatial scales, while they tend to remain unchanged when spatial scales reach certain values.
However, when $b_{1}=0.5\omega/2\tau$ and $b_{2}=0.5\omega/2\tau^{2}$, and $b_{1}=0.9\omega/2\tau$ and $b_{2}=0.9\omega/2\tau^{2}$,
it remains almost unchanged as the spatial scale increases.
This is because for zero and small chirp, where the energy supplied by the external field is relatively low,
the energy of system is mainly derived from the work done by the external field.
Specifically, the work done increases as the spatial scale increases and reaches saturation when the spatial scale reaches a certain value, which in turn causes the particle number to remain unchanged.
In contrast, for moderate and large chirp, the impact of the external field spatial scale is rather weak,
the energy of the system derives from the introduction of external field chirp.
Thus the total particle number exhibits a relatively flat variation with the increase of spatial scales.
Meanwhile, the total particle number with the quadratically symmetric chirped field reaches a higher stable value compared to that of the quadratically asymmetrically one,
see Figs. \ref{fig:3}(a) and \ref{fig:3}(b).      

\begin{table}[htbp]  
	\centering  
	\caption{Total number of the created particles in the quadratically symmetric and quadratically asymmetric high-frequency chirped fields, and the ratio of them. The upper and lower parts of the table represent different chirp for $\lambda=10\ \mathrm{m}^{-1}$ and different spatial scales for $b_{1}=0.5\omega/2\tau$ and $b_{2}=0.5\omega/2\tau^{2}$, respectively. The field parameters are the same as in Fig. \ref{fig:1}.}    
	\begin{tabular}{c c c c c c}  
		\hline
		\hline
		$\lambda$ & $b_{1}$ & $b_{2}$ & $\bar{N}_{\mathrm{sym}}$ & $\bar{N}_{\mathrm{asym}}$ & $\bar{N}_{\mathrm{sym}}/\bar{N}_{\mathrm{asym}}$ \\
		\hline
		& $0.1\omega/2\tau$ & $0.1\omega/2\tau^{2}$ & 0.292 & 0.185 & 1.578 \\
		$10\ \mathrm{m}^{-1}$ & $0.5\omega/2\tau$ & $0.5\omega/2\tau^{2}$ & 3.374 & 1.928 & 1.750 \\
		& $0.9\omega/2\tau$ & $0.9\omega/2\tau^{2}$ & 5.814 & 3.666 & 1.586 \\
		\hline
		$b_{1}$ & $b_{2}$ & $\lambda$ & $\bar{N}_{\mathrm{sym}}$ & $\bar{N}_{\mathrm{asym}}$ & $\bar{N}_{\mathrm{sym}}/\bar{N}_{\mathrm{asym}}$ \\
		\hline
		& & $2\ \mathrm{m}^{-1}$ & 3.407 & 1.901 & 1.792 \\
		$0.5\omega/2\tau$ & $0.5\omega/2\tau^{2}$ & $10\ \mathrm{m}^{-1}$ & 3.374 & 1.928 & 1.750 \\
		& & $300\ \mathrm{m}^{-1}$ & 2.513 & 1.774 & 1.417 \\
		\hline
		\hline
	\end{tabular}   
	\label{tab:I}      
\end{table}        

It is seen that the total particle number of the quadratically symmetric chirp is nearly twice that of the quadratically asymmetric chirp.
For the case of $\lambda=10\ \mathrm{m}^{-1}$, when $b_{1}=0.5\omega/2\tau$ and $b_{2}=0.5\omega/2\tau^{2}$, the maximum ratio between
them can reach 1.750, and for the case of $b_{1}=0.5\omega/2\tau$ and $b_{2}=0.5\omega/2\tau^{2}$, when $\lambda=2\ \mathrm{m}^{-1}$, the maximum ratio of them can reach 1.792.

\subsection{Low frequency field}  
\subsubsection{Momentum distribution of created particles}    

\begin{figure}[htbp]
	\centering
	\includegraphics[width=1\textwidth,height=0.2\textheight]{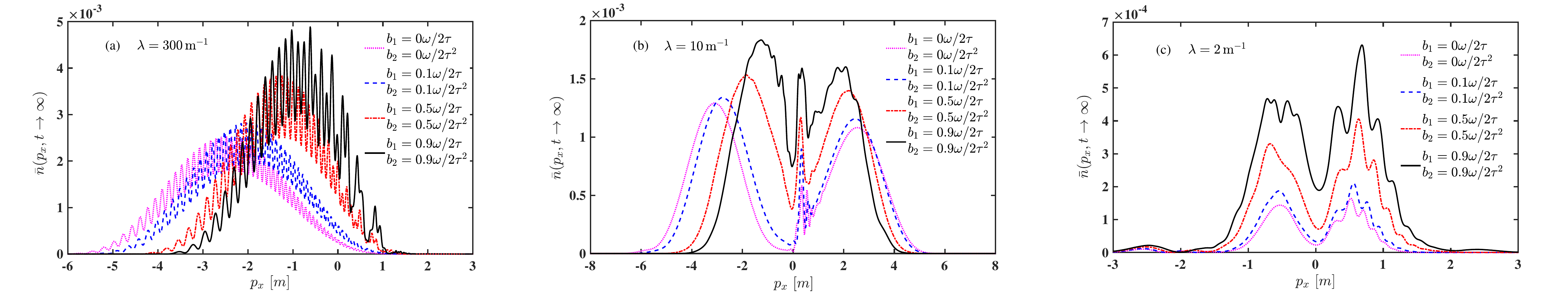}  
	\caption{Reduced momentum spectrum with different quadratically symmetric chirp for various spatial scales in the low-frequency field: (a), (b), and (c) represent the cases where $\lambda=300\ \mathrm{m}^{-1}$, $\lambda=10\ \mathrm{m}^{-1}$, and $\lambda=2\ \mathrm{m}^{-1}$, respectively.
		The field parameters are $\epsilon=0.5$, $\omega=0.1\ \mathrm{m}$, $\tau=25\ \mathrm{m}^{-1}$, and $\phi=\pi/2$.}
	\label{fig:4}    
\end{figure}

The effects of the quadratically symmetric chirp on the momentum spectrum with different spatial scales in the low-frequency field  
are shown in Fig. \ref{fig:4}.        
Note that the momentum spectrum under the quadratically symmetric chirped field for
$\lambda=300\ \mathrm{m}^{-1}$, $\lambda=10\ \mathrm{m}^{-1}$, and $\lambda=2\ \mathrm{m}^{-1}$ at zero chirp are the same as the case of Fig. 2(a) in Ref. \cite{Osman2023}.    

As seen from Fig. \ref{fig:4}(a),     
at the large spatial scale $\lambda=300\ \mathrm{m}^{-1}$,
the momentum spectrum exhibits a bell-shaped structure, accompanied by the obvious oscillation.
Furthermore, it can be observed that the momentum spectrum shifts towards the positive direction.
The WKB approximation in Sec. \ref{section4} can be applied to further qualitatively explain the oscillations and the shifts of the momentum spectrum.   

When the spatial scale is reduced to $\lambda=10\ \mathrm{m}^{-1}$ and the narrow spatial scale $\lambda=2\ \mathrm{m}^{-1}$,
as seen from Figs. \ref{fig:4}(b) and \ref{fig:4}(c),             
the overall characteristic of the momentum spectrum shows a peak-valley-peak structure at the middle.
The two main peaks occur because the ponderomotive forces introduced by the finite spatial scale of the external field push the pairs apart, causing them to accelerate in the $x$ and $-x$ directions, respectively.
And the valley at $p=0$ results from the symmetry of applied electric field in $\pm x$ directions,
where the particle pairs accelerating in the positive or negative directions cancel each other out.  
Note that in Figs.  \ref{fig:4}(b) and \ref{fig:4}(c),   
when the spatial scale decreases from $\lambda=10\ \mathrm{m}^{-1}$ to $\lambda=2\ \mathrm{m}^{-1}$,  
the oscillations near the valley will extend toward the peak of the positive direction, i.e.,
at $\lambda=2\ \mathrm{m}^{-1}$, the positive direction peak of the momentum spectrum will be replaced by weak oscillations.
In addition, the central valley will become broader, which is caused by the stronger ponderomotive force at narrow spatial scales. Particularly, the very small peaks can be seen on both sides of the momentum spectrum,
which is caused by interference effects between pairs.  

\begin{figure}[htbp]
	\centering
	\includegraphics[width=1\textwidth,height=0.15\textheight]{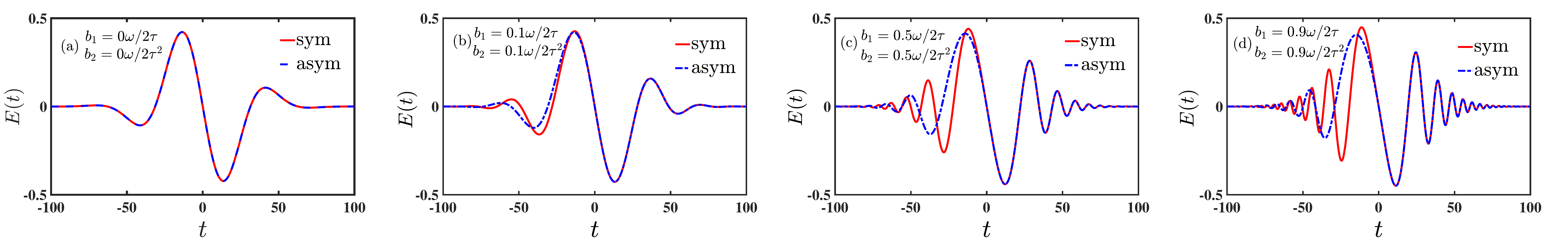}       
	\caption{The temporal variation of the electric field for different quadratically symmetric and quadratically asymmetric chirped fields,
		(a)--(d) represent the cases of $b_{1}=b_{2}=0$, $b_{1}=0.1\omega/2\tau$ and $b_{2}=0.1\omega/2\tau^{2}$, $b_{1}=0.5\omega/2\tau$ and $b_{2}=0.5\omega/2\tau^{2}$, and $b_{1}=0.9\omega/2\tau$ and $b_{2}=0.9\omega/2\tau^{2}$, respectively.
		The field parameters are the same as in Fig. \ref{fig:4}.}   
	\label{fig:5}          
\end{figure}

Meanwhile, as shown in Fig. \ref{fig:4}, the peaks of the momentum spectrum increase and the oscillations intensify as the chirp increases.    
The increase of the momentum spectrum peak with increasing chirp is attributed to the higher effective frequency,
which leads to the elevation of the system energy.
Subsequently, the reason why the oscillation becomes more intense as the chirp increases can be explained by referring to the Keldysh parameter.
From the given parameters in this paper, it can be known that $\zeta\ll1$,  
so the pairs production occurs predominantly by tunneling process.
Moreover, it can be seen from the expression of Keldysh that the tunneling effect is so highly sensitive to the electric field strength that it becomes more pronounced as the field strength increases.
We present the variation of the time-dependent field strength with the chirp, as shown by the red solid line in
Fig. \ref{fig:5}.   

It can be observed that the main large peak value of the electric field increases and the number of the large peaks also increases with the increase of chirp.
It is note that pairs are predominantly created at the large peaks values of electric field and their vicinity \cite{Ababekri2020}.
Therefore, the number of pairs generated through the tunneling process increases in this situation, thereby enhancing the  interference effect between the pairs, i.e.,
the oscillations in the momentum spectrum strengthen with increasing chirp.
Moreover, it can be clearly that the peak values of the momentum spectrum drop significantly when the spatial scale of the external field decreases, which is due to the decreased work done by the external electric field and the consequent reduction in the energy of system.   

\begin{figure}[htbp]      
	\centering
	\includegraphics[width=0.8\textwidth,height=0.4\textheight] {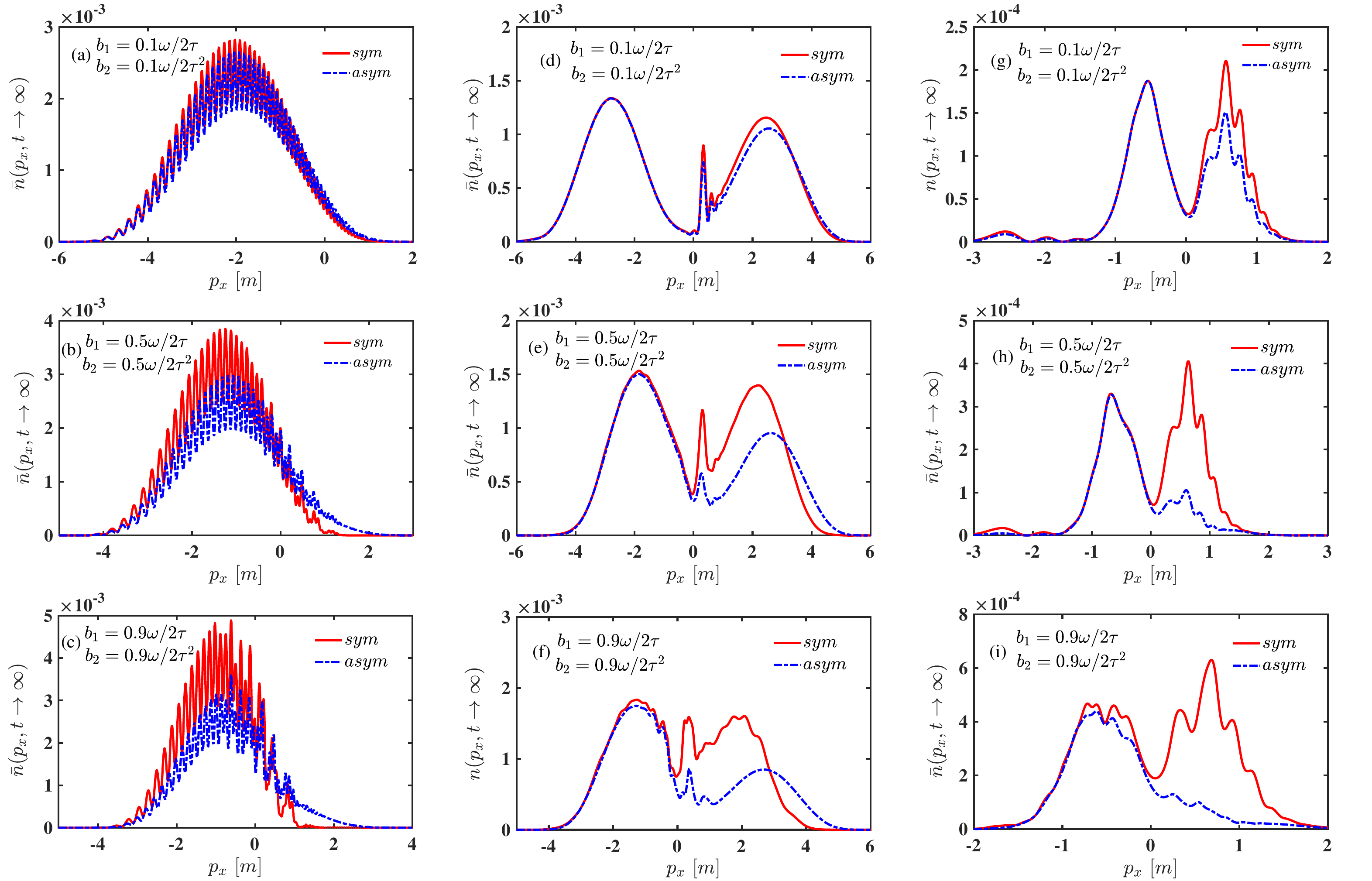}  
	\caption{Comparison of the reduced momentum spectrum with quadratically symmetric and quadratically asymmetric chirp in the low-frequency field.
		From left to right, the columns (a)--(c), (d)--(f), and (g)--(i) represent the cases of $\lambda=300\ \mathrm{m}^{-1}$, $\lambda=10\ \mathrm{m}^{-1}$, and $\lambda=2\ \mathrm{m}^{-1}$, respectively.
		The field parameters are the same as in Fig. \ref{fig:4}.}
	\label{fig:6}
\end{figure}

Furthermore, a comparison of the momentum spectrum with quadratically symmetric and quadratically asymmetric chirped fields is presented in Fig. \ref{fig:6}.   

It is seen that the momentum spectrum for the quadratically symmetric chirped field exhibits the stronger oscillations and the higher peak values than that for the quadratically asymmetric one.
The stronger oscillations can be analyzed by comparing the time-dependent electric field of the quadratically symmetric and quadratically asymmetric chirp, as shown in Fig. \ref{fig:5}.      
It can be observed that
in the $+t$ direction, the variations of the time-dependent electric field of the quadratically symmetric and quadratically asymmetric chirp are the same,
while in the $-t$ direction, the main peak value and the number of these large peaks of the time-dependent electric field
in the quadratically symmetric chirped field are higher and larger than those for the quadratically asymmetric one,
which leads to the increased number of pairs in the quadratically symmetric chirped field, then enhances the interference effect between pairs.
Subsequently, the higher peak values can be attributed to energy and ponderomotive force.
At $\lambda=300\ \mathrm{m}^{-1}$, the peak enhancement is mainly caused by the increase of the total energy of the system, while at $\lambda=10\ \mathrm{m}^{-1}$ and $\lambda=2\ \mathrm{m}^{-1}$,
the peak enhancements result from both the increase of energy and the increase of ponderomotive force.

\subsubsection{Total number of created particles}

\begin{figure}[htbp]
	\includegraphics[width=0.8\textwidth,height=0.35\textheight]{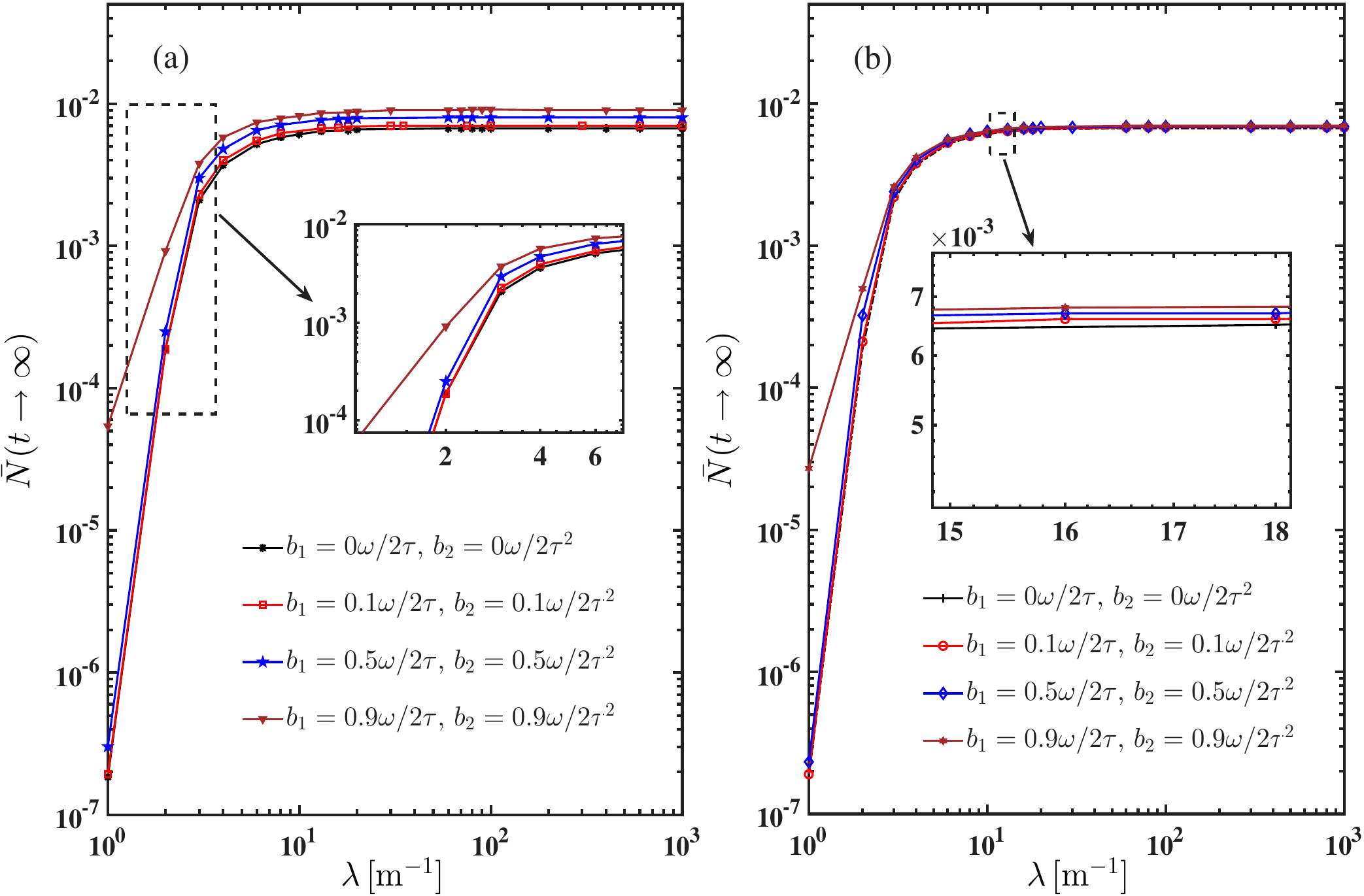}     
	\caption{The variations of the total number of the created particles with spatial scales for different quadratically symmetric and quadratically asymmetric chirp in the low-frequency field, (a) and (b) represent the cases of the quadratically symmetric and quadratically asymmetric chirp, respectively.
		The inset in (a) shows the variation of the total number of particles with quadratically symmetric chirp in small spatial regions, and the inset in (b) shows the variation of the total number of particles that reach a steady value with quadratically asymmetric chirp.
		The field parameters are the same as in Fig. \ref{fig:4}.}       
	\label{fig:7}      
\end{figure}

The variations of the total number of created particles with spatial scales for different quadratically symmetric and quadratically asymmetric chirp in the low-frequency field are shown in Fig. \ref{fig:7}. 
Note that Fig. \ref{fig:7}(b) is the same case of Fig. 3(b) in Ref. \cite{Osman2023}.     

It is seen that the total number of the created particles in quadratically symmetric and quadratically asymmetric chirped fields both increase with increasing spatial scales, and they remain constant when the spatial scales reach a certain value.
This can be understood through the work done by the external field, which increases as the spatial scale increases,  
resulting in an increase in the energy of system.
However, when the spatial scale reaches a certain value, the work done by the external field reaches saturation,
so the generated pairs remain unchanged.
Particularly, the reached stable value of the total number of particles in the quadratically symmetric and quadratically asymmetric chirped fields both increase with chirp,
and their values in the quadratically symmetric chirped field are significantly higher than those in the quadratically asymmetric  
one.

To illustrate the increase of the total particle number in the quadratically symmetric chirped field,
two representative examples with $\lambda=2\ \mathrm{m}^{-1}$, and $b_{1}=0.9\omega/2\tau$ and $b_{2}=0.9\omega/2\tau^{2}$ are taken.    The total particle number of the quadratically symmetric and quadratically asymmetric chirped field, and the ratio of them are shown in Table \ref{tab:II}.    
It is seen that the total particle number of the quadratically symmetric chirp is nearly twice that of the quadratically asymmetric chirp.
The maximum increase occurs at $\lambda=2\ \mathrm{m}^{-1}$, $b_{1}=0.9\omega/2\tau$ and $b_{2}=0.9\omega/2\tau^{2}$, reaching 1.840 times.     
\begin{table}[htbp]
	\centering
	\caption{Total number of the created particles in the quadratically symmetric and quadratically asymmetric low-frequency chirped fields, and the ratio of them.
The upper and lower parts of the table represent different chirp for $\lambda=2\ \mathrm{m}^{-1}$ and different spatial scales for $b_{1}=0.9\omega/2\tau$ and $b_{2}=0.9\omega/2\tau^{2}$, respectively.
  The field parameters are the same as in Fig. \ref{fig:4}.}    
 	\begin{tabular}{c c c c c c}     
		\hline
	    \hline
			 $\lambda$  &  $b_{1}$ & $b_{2}$ & $\bar{N}_{sym}$ & $\bar{N}_{asym}$ &   $\bar{N}_{sym}/\bar{N}_{asym}$  \\
			\hline
			& $0.1\omega/2\tau$ & $0.1\omega/2\tau^{2}$ & $2.5\times10^{-4}$ & $2.1\times10^{-4}$ & 1.190  \\
	   $2\ \mathrm{m}^{-1}$   &   $0.5\omega/2\tau$ & $0.5\omega/2\tau^{2}$ & $5.3\times10^{-4}$ & $3.2\times10^{-4}$ & 1.656 \\
			&  $0.9\omega/2\tau$ & $0.9\omega/2\tau^{2}$ & $9.2\times10^{-4}$ & $5.0\times10^{-4}$ & 1.840 \\
			\hline
			$b_{1}$ & $b_{2}$ &  $\lambda$ & $\bar{N}_{sym}$  & $\bar{N}_{asym}$ &  $\bar{N}_{sym}/\bar{N}_{asym}$   \\
			\hline
			 &    &   $2\ \mathrm{m}^{-1}$  &  $2.5\times10^{-4}$  &  $2.1\times10^{-4}$                               & 1.840\\
			$0.9\omega/2\tau$ & $0.9\omega/2\tau^{2}$ & $10\ \mathrm{m}^{-1}$ & $6.5\times10^{-3}$ & $6.2\times10^{-3}$ & 1.281 \\
			&   & $300\ \mathrm{m}^{-1}$ &  $7.0\times10^{-3}$ & $6.8\times10^{-3}$                                        & 1.286 \\
			\hline
			\hline
		\end{tabular}
	  \label{tab:II}  
   \end{table}  

\section{TURNING-POINT ANALYSIS}       \label{section4}

In order to clearly understand the features of the obtained momentum spectrum in the low-frequency field, such as the oscillations and the shifts,
this section discusses the $e^{+}e^{-}$ pair production using the WKB approximation
\cite{Hebenstreit2009PRL}.
The pair creation from vacuum in the spatially inhomogeneous electric fields with quadratically symmetric chirp
is similar to the one-dimensional over-the-barrier scattering problem in quantum mechanics
\cite{Brezin1970,Dunne2009,Marinov1977}.
Thus, it can be expressed as an effective ``Schr{\"o}dinger-like equation'', and it is given by
\cite{Heading1962,Gong2020}
\begin{equation}
\frac{\mathrm{d}^{2}\varphi(\mathbf{p},t)}{\mathrm{d}t^{2}}+ \Theta^{2}(\mathbf{p},t)\varphi(\mathbf{p},t)=0,
\end{equation}
where $\Theta(\mathbf{p},t)=\sqrt{m^{2}+p_{\perp}^{2}+\left[p_{x}-e A(t)\right]^{2}}$,
$p_{\perp}$ and $p_{x}$ are the momenta perpendicular and parallel to the external field, respectively,
and the incident energy is $m^{2}+p_{\perp}^{2}$.

This can be compared to the ``Schr{\"o}dinger equation'' \cite{Landau2003}, where the effective scattering potential described by the equation is given by $V(t)=-\left[p_{x}-eA(t)\right]^{2}$.
It indicates that the scattering potential $V(t)$ changes dramatically as $p_{x}$ varies, which leads to the scattering resonances.
The momentum distribution function of the created pairs involving multiple turning points
is obtained by the reflection coefficient, and it is given by
\cite{Dumlu2011,Hebenstreit2009PRL,Froman1970}
\begin{equation}
f(\mathbf{p}) \approx \sum_{t_p}\exp(-2 K_{\mathbf{p}}^p)
+ \sum_{t_p \neq t_{p'}} 2(-1)^{p-p'}\cos \left( 2 \theta_{\mathbf{p}}^{(p,p')} \right)\exp(-K_{\mathbf{p}}^p - K_{\mathbf{p}}^{p'}),
\label{eq:21}
\end{equation}
with \[K_{\mathbf{p}}^p = \left| \int_{t_p^*}^{t_p} \Theta(\mathbf{p}, t) \, \mathrm{d}t \right|, \quad
\theta_{\mathbf{p}}^{(p,p')}=\int_{\text{Re}(t_{p})}^{\text{Re}(t_{p'})} \Theta(\mathbf{p}, t) \, \mathrm{d}t,\]
where $t_p$ and $t_{p'}$ represent different pairs of turning points, which are the zeros of $\Theta(\mathbf{p},t)$,
and the pairs of turning points closest to the real axis tend to dominate in the semiclassical regime.
The first term of Eq. (\ref{eq:21}) is an exponentially decaying distribution, and the second term represents an interference term accompanied by exponentially decaying terms.
It is clear that $K_{\mathbf{p}}^p$ and     $\theta_{\mathbf{p}}^{(p,p')}$ determine the number of created pairs and the intensity of interference effects.
Therefore, the oscillation behavior of the momentum spectrum can be understood in terms of interference effects between pairs of
turning points
and the shift of the momentum spectrum can be explained by analyzing the pair of turning points closest to the real axis
\cite{Dumlu2010,Dumlu2010a}.

\begin{figure}[htbp]
	\centering
	\includegraphics[width=0.6\textwidth,height=0.2\textheight]{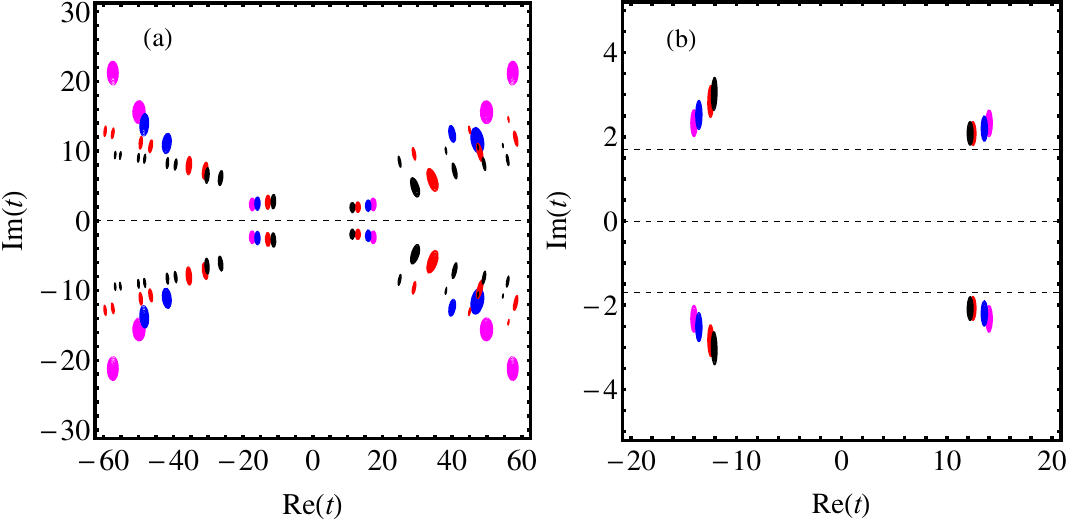}     
	\caption{Contour plots of $\left| \Theta (\mathbf{p},t) \right|^2$ in the complex $t$ plane,
		showing the location of turning points for $p_{\bot}=0$ of the low-frequency field, where $\Theta(\mathbf{p},t)=0$ .
		These plots are for the momentum $p_{x}=-1\ \mathrm{m}$ in panel (a), whereas the momentum is , $p_{x}=-2\ \mathrm{m}$, $p_{x}=-1.2\ \mathrm{m}$, and $p_{x}=-0.6\ \mathrm{m}$ in panel (b).
		Both panels present turning points for four sets of chirp parameters, $b_{1}=b_{2}=0$  (magenta), $b_{1}=0.1\omega/2\tau$,   $b_{2}=0.1\omega/2\tau^{2}$ (blue), $b_{1}=0.5\omega/2\tau$, $b_{2}=0.5\omega/2\tau^{2}$ (red), and   $b_{1}=0.9\omega/2\tau$ and $b_{2}=0.9\omega/2\tau^{2}$ (black). The field parameters are the same as in Fig. \ref{fig:4}.}    
	\label{fig:8}    
\end{figure}

The structure of the turning points for different quadratically symmetric chirp at the large spatial scales in the low-frequency field are shown in Fig. \ref{fig:8}.
Note that $p_x = -1\ \mathrm{m}$ is taken in Fig. \ref{fig:8}(a), which corresponds to the region of the strongest oscillation in the momentum spectrum of Fig. \ref{fig:4}(a),         
and the value of $p_x$ in Fig. \ref{fig:8}(b) is taken at the position of the large peak in the momentum spectrum of Fig. \ref{fig:4}(a).    
That is, $p_x=-2.4\ \mathrm{m}$ for $b_{1}=b_{2}=0$ and $p_x=-2\ \mathrm{m}$ for
$b_{1}=0.1\omega/2\tau$ and $b_{2}=0.1\omega/2\tau^{2}$, $p_x=-1.2\ \mathrm{m}$ for $b_{1}=0.5\omega/2\tau$ and $b_{2}=0.5\omega/2\tau^{2}$, and  $p_x=-0.6\ \mathrm{m}$ for $b_{1}=0.9\omega/2\tau$ and $b_{2}=0.9\omega/2\tau^{2}$, respectively.  

It can be observed that the structure of the turning points take on a saddle-shaped distribution from Fig. \ref{fig:8}(a),         
and there are two central pairs of turning points that are approximately equidistant from the real axis, this shows us that interference effects will be pronounced, and then Eq. (\ref{eq:21}) can be simply expressed as    
\begin{equation}   
f(\mathbf{p}) \approx \exp \left(-2 K_{\mathbf{p}}^{(1)}\right) + \exp \left( -2 K_{\mathbf{p}}^{(2)}\right)-
2\cos\left(2\theta_{\mathbf{p}}^{(1,2)}\right) \exp\left(-K_{\mathbf{p}}^{(1)}- K_{\mathbf{p}}^{(2)} \right),
\end{equation}
with \[ \theta_{\mathbf{p}}^{(1,2)}=\int_{\text{Re}(t_{1})}^{\text{Re}(t_{2})} \Theta(\mathbf{p}, t) \, \mathrm{d}t.\]

As can be seen in Fig. \ref{fig:4}(a), the oscillations can be observed, which intensifies as the chirp increases.      
This is illustrated by the two central turning points pairs shown in Fig. \ref{fig:8}(a), where it can be observed that the distance of the two central pairs from the imaginary axis gets closer and closer as the chirp increases.     
The distribution of turning points for the four sets of chirp parameters and momentum values are shown in Fig. \ref{fig:8}(b). It can be seen that the turning points from magenta to black are getting closer and closer to the real axis. This means that the particles creation rate in the case of $b_{1}=0.9\omega/2\tau$ and    $b_{2}=0.9\omega/2\tau^{2}$ is large comparable to the others. and the location of peak value is shift from $p_{x}=-2.4\ \mathrm{m}$  to   $p_{x}=-0.6\ \mathrm{m}$.
This precisely corresponds to the shifts of the momentum spectra for $b_{1}=b_{2}=0$, $b_{1}=0.1\omega/2\tau$ and   $b_{2}=0.1\omega/2\tau^{2}$, $b_{1}=0.5\omega/2\tau$ and $b_{2}=0.5\omega/2\tau^{2}$, and $b_{1}=0.9\omega/2\tau$ and $b_{2}=0.9\omega/2\tau^{2}$ in Fig. \ref{fig:4}(a).      
However, the positions of the pair of turning points in black are almost the same as those of the pair of turning points in red,  
which explains that the momentum spectrum shift from $b_{1}=0.5\omega/2\tau$ and $b_{2}=0.5\omega/2\tau^{2}$ to       $b_{1}=0.9\omega/2\tau$ and $b_{2}=0.9\omega/2\tau^{2}$ is not obvious.  

\section{CONCLUSIONS AND OUTLOOK}       \label{section5}

The momentum spectrum and the total number of the created particles are studied in the spatially inhomogeneous electric
fields with the quadratically symmetric chirp using DHW method in high- and low-frequency fields.
In particular, they are compared with those for the quadratically asymmetric chirped field.
It is found that the momentum spectrum of the created particles shows the stronger oscillations and the higher peaks under
the quadratically symmetric chirped field in both high- and low- frequency fields,
and a significant widening is observed only in the high-frequency field.
It is also found that the total number of created particles in the quadratically symmetric chirped field increases as the chirp increases, and it is nearly twice that of the quadratically asymmetric chirped field.
Furthermore, in the high-frequency field, it is concluded that the $e^{+}e^{-}$ pairs are mainly generated through multiphoton absorption.
However, in the low-frequency field, the pairs are mainly produced through tunneling process.
Finally, the WKB approximation is applied to explain the features of the momentum spectrum,
such as the oscillations and the shifts of the momentum spectrum at large spatial scales in the low-frequency field.
The oscillations in the momentum spectrum can be explained by the pronounced interference effects between two central turning points pairs that are approximately equidistant from the real axis, and
the shifts of the momentum spectrum can be analyzed using the pair of turning points closest to the real axis, where it is shown that this pair gets progressively closer to the real axis from zero chirp to moderate chirp.

This research indicates that quadratically symmetric chirped fields with spatially inhomogeneous plays a crucial role in vacuum $e^{+}e^{-}$ pair production,
and the external field model provides a possibility to achieve high efficiency $e^{+}e^{-}$ pair production.
However, it is only considered that $e^{+}e^{-}$ pair production occurs in the quadratically symmetric chirped electric fields,
and it is believed that the presented research can be extended to quadratically symmetric chirped electromagnetic fields or various combined fields.

\section*{Acknowledgments}

This work is supported by the National Natural Science Foundation of China (NSFC) under Grant Nos. 12565023, 12535015, and 12375240, the Youth Science and technology Foundation of Gansu Province under Grant No. 24JRRA276,
and the Gansu Science and Technology Program under Grant No. 23JRRA681.

\end{document}